\documentclass[twoside,12pt]{article}
\usepackage{epsfig}

\begin{document}

\title{ Distinguishing the $0\nu\beta\beta$-decay
mechanisms }

\author{Fedor \v Simkovic\footnote{{\it On leave from:}
 Department of Nuclear physics, Comenius University, 
SK--842 15 Bratislava, Slovakia} and Amand Faessler\\
Institut f\"ur Theoretische Physik der Universit\"at T\"ubingen\\ 
D-72076 T\"ubingen, Germany}

\maketitle

\begin{abstract} 
Many new $0\nu\beta\beta$-decay experiments are planned
or in preparation. If the $0\nu\beta\beta$-decay will be 
detected, the key issue will be what is the dominant 
mechanism of this process. By measuring only transitions
to the ground state one can not  distinguish among many of
the $0\nu\beta\beta$-decay mechanisms (the light
and heavy Majorana neutrino exchange mechanisms, 
the trilinear R-parity breaking mechanisms etc.).
We  show that if the ratio of the $0\nu\beta\beta$-decay
half-lifes for transitions to  the $0^+$ first excited 
and ground states is determined both theoretically and
experimentally, it might be  possible to determine, which
 $0\nu\beta\beta$-decay mechanisms is dominant. 
For that purpose the corresponding nuclear matrix elements 
have to be evaluated with high reliability.  The present work
is giving strong motivations for experimental studies of
the $0\nu\beta\beta$-decay transitions to the first excited
$0^+$ states of the final nuclei.

PACS number(s): 23.40.Bw, 13.35.Bv, 21.60.Jz, 12.60.Cn

\end{abstract}

\section{Introduction}

Today, there is a strong belief for the existence of new physics
beyond the standard model (SM). Recent Super-Kamiokande \cite{KAM} 
and SNO \cite{sno} results are considered as strong evidence 
in favor of  oscillations of the atmospheric and solar 
neutrinos, i.e. in favor of  neutrino masses and mixing.
By making phenomenologically viable assumptions different
predictions for the mass spectrum of  light neutrinos 
have been proposed \cite{smir,bil99,haug}. 
The problem of the neutrino masses can
not be solved without complementary information from the
side of the neutrinoless double beta decay. 
At present, the lower 
limits on the half-life of the neutrinoless double beta decay
($0\nu\beta\beta$-decay) 
and thus the upper limit of effective electron neutrino
mass $<m_\nu >_{ee}$ allows to
discriminate among different neutrino mixing 
scenarios \cite{smir,bil99,haug}. In these analysis one assumes 
the dominance of the conventional $0\nu\beta\beta$-decay 
mechanism via the exchange of light Majorana neutrinos
between the decaying neutrons. 

The neutrinoless double beta decay 
\cite{doi83,ver86,pv,moh99,fae98},
\begin{equation}
(A,Z) \rightarrow (A,Z+2) + 2e^-,
\label{eq:1}
\end{equation}
converting a nucleus (A,Z) into  (A,Z+2) under emission
of two electrons, violates the lepton number by two units and
is forbidden in the standard model (SM). Any experimental 
evidence for this rare nuclear process will constitute 
evidence for  new physics beyond the SM.

Essential progress in the exploration of the decay
both from theoretical and experimental sides has been achieved
in the last few years.
The considerably improved experimental lower bounds on
the half-life of various isotopes enhance the potential
of testing the different concepts of physics beyond the SM
\cite{bau99,avi,cuore,nemo3}.
The $0\nu\beta\beta$-decay has been found to be a probe of all
kinds of new physics beyond the SM. Indeed the existing lower
limits on this process imply very stringent constraints on
new physics scenarios such as the left-right symmetric models,
supersymmetric (SUSY) models with R-parity violation and many others.
The Grand unified theories (GUT) and R-parity violating SUSY
models offer a plethora of the $0\nu\beta\beta$-decay mechanisms 
triggered by exchange of neutrinos, neutralinos, gluinos,
leptoquarks, etc. If the $0\nu\beta\beta$-decay will be observed in 
one of the current or planned experiments, the main issue will
be what is the dominant mechanism of this process.  

It has been shown by Doi et al. \cite{doi83} that it is possible to distinguish
the light and heavy Majorana neutrino  mass mechanisms of the
$0\nu\beta\beta$-decay from those
induced by the right-handed currents by measuring the differential
characteristics as well as the transitions to the $2^+$ excited state.
However, there is a class of $0\nu\beta\beta$-decay mechanisms,
which one cannot distinguish  from each other kinematically.
The light and heavy Majorana neutrino
mass and the trilinear R-parity breaking mechanisms
constitute such a group. In this contribution we shall discuss 
in details the possibility to distinguish among them experimentally.
We shall show that it is possible to do it
by measuring both the transitions to the first 
excited $0^+_1$ and to the $0^+_{g.s.}$ ground  states. 
We note that by measuring only $0\nu\beta\beta$-decay
transitions to the ground state one can not  distinguish among these
mechanisms even in the case the experimental results for more isotopes
are known as the ratios of  nuclear matrix elements
for different nuclei associated with
different  mechanisms do not differ by each from
other \cite{si99}. 

\section{Distinguishing between
 the $0\nu\beta\beta$-decay mechanisms}

The half-life of the $0\nu\beta\beta$-decay to a  $0^+$ final  state 
associated with the above mentioned mechanisms  is partly dictated 
by the nuclear matrix element ${\cal M}_{\cal{K}}$, 
partly by the phase space factor $G_{01}(0^+)$ and
partly by the  lepton number violating parameter $\eta_{\cal{K}}$. 
In our analysis we shall assume that one mechanism of the
$0\nu\beta\beta$-decay at a time dominates. Then the inverse half-life of
the $0\nu\beta\beta$-decay  takes the form \cite{doi83,fae98}
\begin{equation}
[T_{1/2}^{0\nu}(0^+_{g.s.}\rightarrow 0^+)]^{-1} = \left|\eta_{\kappa}\right|^2~ 
|{\cal M}_{\kappa}(0^+)|^2~G_{01}(0^+). 
\label{eq:2}   
\end{equation}
Here, the index $\kappa$  denotes one of the mechanisms of 
the $0\nu\beta\beta$-decay, which will be discussed in the subsequent sections.

\begin{figure}[tb]
\epsfxsize=8.0cm
\begin{center}
\epsfig{file=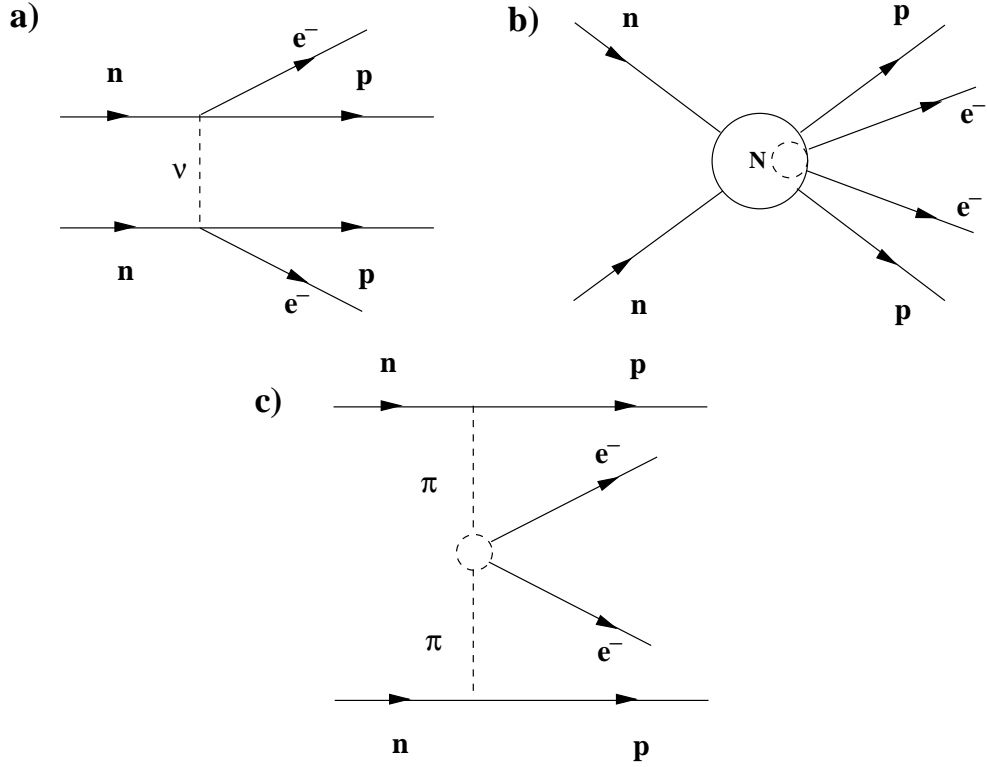,scale=0.5}
\end{center}
\caption{
The two nucleon exchange potential in the case of the light neutrino-exchange (a), heavy neutrino-exchange
and R-parity breaking SUSY mechanisms of the $0\nu\beta\beta$-decay.
\label{fig.1}}
\end{figure}

The ratio $\xi_{full}$ of the $0\nu\beta\beta$-decay half-lifes
for the transition to the first excited $0^+_1$  and to 
the ground  state is free of unknown parameters from
the particle physics side. We have
\begin{eqnarray}
\xi_{full} = \frac{T^{0\nu}_{1/2}(0^+_{g.s.}\rightarrow 0^+_1)}
{T^{0\nu}_{1/2}(0^+_{g.s.}\rightarrow 0^+_{g.s.})}~=~
\frac{G_{01}(0^+_{g.s.})}{G_{01}(0^+_{1})} ~
\Big(\frac{{\cal M} (0^+_{g.s.})}{{\cal M}(0^+_{1})}\Big)^2 = \xi_{kin.}~ \xi_{\cal M}.
\label{eq:3}
\end{eqnarray}
We see that the  ratio $\xi_{full}$
can be factorized as product of two 
factors,  $\xi_{kin.}$ and $\xi_{\cal M}$.
The factor $\xi_{kin.}$, which is given as ratio of
phase space integrals for transitions to ground and
excited  states, is larger than unity due to the
smaller Q value for the transition to the excited state, 
where the available phase space is reduced.
We note that this factor is the same for all  
$0\nu\beta\beta$-decay mechanisms considered. On the other side,
the  squared ratio of the corresponding nuclear matrix elements
$\xi_{\cal M}$ is strongly depending on the specific 
$0\nu\beta\beta$-decay mechanism.  The 
two-nucleon exchange potentials associated with the light
and heavy Majorana neutrino mass mechanisms as well as
with the trilinear R-parity breaking SUSY mechanism
of the $0\nu\beta\beta$-decay differ considerably by each of other
(see Fig. \ref{fig.1}). Their form will be given in the forthcoming sections. 

From the above discussion, we therefore conclude that
 the  ratio $\xi_{full}$ for a given isotope
depends on the specific $0\nu\beta\beta$-decay mechanism. 
Its value can be determined both theoretically and experimentally.
The theoretical study requires  the evaluation of
kinematical factors and the reliable calculation of 
nuclear matrix elements involved for different $0\nu\beta\beta$-decay 
mechanisms \cite{suhi,sim01}. 
The experimental determination of $\xi_{full}$ is 
difficult as it requires the observation of   transition
to the $0^+$ excited and to the ground state. However,  its measurement
makes it possible  to decide, which of the $0\nu\beta\beta$-decay
mechanisms is the dominant one.

\section{Nuclear structure details }

The standard proton-neutron Quasiparticle Random Phase Approximation
(pn-QRPA), which is based on the quasiboson approximation, and
its renormalized version pn-RQRPA, which includes anharmonicities,
have been  choice in the
calculation of the double beta decay matrix elements \cite{simn96,zare}. 
Within these nuclear structure approaches the matrix element 
${\cal M}_{(i)}(0^+)$ entering the Eq. (\ref{eq:2}) can be written as
\begin{eqnarray}
\label{eq:4}
{\cal M}_{\kappa} & = &
\sum_{J^{\pi}} \sum_{{{p n p' n' } \atop m_i m_f {\cal J}  } }
(-)^{j_{n}+j_{p'}+J+{\cal J}}(2{\cal J}+1)
\left\{\matrix{
j_p & j_n & J \cr
j_{n'} & j_{p'} & {\cal J}}\right\} \nonumber \\
&& \times
\langle p(1), p'(2);{\cal J} | f(r_{12}) \tau_1^+ \tau_2^+ 
{\cal O}_{\kappa}  (12)
f(r_{12}) |n(1) ,n'(2);{\cal J}\rangle \nonumber \\
&& \times
\langle 0_f^+ \parallel
\widetilde{[c^+_{p'}{\tilde{c}}_{n'}]_J} \parallel J^\pi m_f\rangle
\langle J^\pi m_f|J^\pi m_i\rangle
\langle J^\pi m_i \parallel [c^+_{p}{\tilde{c}}_{n}]_J \parallel
0^+_i\rangle.
\end{eqnarray}
Here, $f(r_{12})$ is the short-range correlation function  \cite{fae98}
and ${\cal O}_{\cal K}(12)$ 
represents the coordinate and spin dependent part of the
two-body transition operator. 
 Separating the Fermi (F), Gamow-Teller (GT) and
the tensor (T) contributions we obtain:
\begin{equation}
{\cal O}_{\kappa} (12) = 
- \frac{H_{F}^{\kappa} (r_{12})}{g^2_A}   + 
H_{GT}^{\kappa} (r_{12}) {\bf \sigma}_{12} +
H_{T}^{\kappa} (r_{12}) {\bf S}_{12}.
\label{eq:5}
\end{equation}
The following notations have been used:
\begin{eqnarray}
{\bf r}_{12} &= &{\bf r}_1-{\bf r}_2, ~~~
r_{12} = |{\bf r}_{12}|, ~~~
\hat{{\bf r}}_{12} = \frac{{\bf r}_{12}}{r_{12}},~\nonumber \\
S_{12} &=& 3({\vec{ \sigma}}_1\cdot \hat{{\bf r}}_{12})
       ({\vec{\sigma}}_2 \cdot \hat{{\bf r}}_{12})
      - \sigma_{12}, ~~~ \sigma_{12}=
{\vec{ \sigma}}_1\cdot {\vec{ \sigma}}_2.
\label{eq:6}   
\end{eqnarray}
where ${\bf r}_1$ and ${\bf r}_2$ are coordinates of the beta decaying nucleons.
The explicit form of the radial part of the two-nucleon exchange potential 
for the $0\nu\beta\beta$-decay mechanisms $\kappa$ of interest 
shall be present in the following sections.

The one--body transition densities 
$<J^\pi m_i\parallel [c^+_{p}{\tilde{c}}_{n}]_J\parallel 0^+_i>$,
$<0_f^+\parallel\widetilde{ [c^+_{p}{\tilde{c}}_{n}]_J}\parallel J^\pi m_f> $
and the overlap factor $\langle J^\pi m_f|J^\pi m_i\rangle$ \cite{simr88}
entering Eq. (\ref{eq:5}) must be computed in a nuclear model.
 
In this contribution we shall consider transitions to the first excited 
two-quadrupole phonon $0^+$ state, which can be described as follows:
\begin{equation}
|0^+_1 \rangle = \frac{1}{\sqrt{2}} 
\{\Gamma^{1\dagger}_2 \otimes \Gamma^{1\dagger}_2\}^0
|0^+_{g.s.} \rangle, 
\label{eq:7}
\end{equation}
where $\Gamma^{1\dagger}_2$ is the quadrupole phonon creation operator
for the first excited $2^+$ state of the daughter nucleus. This state is 
defined by
\begin{equation}
|2^+_1 \rangle = \Gamma^{1\dagger}_{2M^+}|0^+_{f}\rangle
 \qquad \mbox{with} \qquad
\Gamma^{1}_{2M^+}|0^+_{f}\rangle=0.
\label{eq:8}
\end{equation}
Here $|0^+_{f}\rangle$ is the  ground state of the final
nucleus.

In a similar way 
the $m^{th}$ excited state of the intermediate odd-odd nucleus
(A,Z+1) is defined, with the angular momentum $J$ and the projection $M$. 
We generate it  by applying the phonon-operator $Q^{m\dagger}_{JM^\pi_{i,f}}$
on the vacuum state $|0^+_{i,f}\rangle$:
\begin{equation}
|m_{i,f}, JM^\pi \rangle = Q^{m\dagger}_{JM^\pi_{i,f}}|0^+_{i,f}\rangle
 \qquad \mbox{with} \qquad
Q^{m}_{JM^\pi_{i,f}}|0^+_{i,f}\rangle=0.
\label{eq:9}
\end{equation}
Here, the indices $i$ and $f$ are defined in respect to the initial
(A,Z) and final (A,Z+2) nuclei, respectively.  

In the framework of the pn-RQRPA the reduced one-body transition densities
for the ground state transition are
\begin{eqnarray}
\label{eq:10}   
<J^\pi m_i\parallel [c^+_{p}{\tilde{c}}_{n}]_J \parallel 0^+_i>
 &=& \sqrt{2J+1} 
(u_{p}^{(i)} v_{n}^{(i)} {\overline{X}}^{m_i}_{(pn, J^\pi)}
+v_{p}^{(i)} u_{n}^{(i)} {\overline{Y}}^{m_i}_{(pn, J^\pi)})
\sqrt{{\cal D}^{(i)}_{pn}},  \\
<0_f^+\parallel\widetilde{ [c^+_{p}{\tilde{c}}_{n}]_J}
\parallel J^\pi m_f> 
&=& \sqrt{2J+1} 
(v_{p}^{(f)} u_{n}^{(f)} 
{\overline{X}}^{m_f}_{(pn, J^\pi)}
+u_{p}^{(f)} v_{n}^{(f)} 
{\overline{Y}}^{m_f}_{(pn, J^\pi)})
\sqrt{{\cal D}^{(f)}_{pn}},
\label{eq:11}   
\end{eqnarray}
where ${\overline{X}}^{m_{i,f}}_{(pn, J^\pi)}$ and
${\overline{X}}^{m_{i,f}}_{(pn, J^\pi)}$ are the forward
and backward-going amplitudes related to the phonon
operator $Q^{m\dagger}_{JM^\pi_{i,f}}$ in Eq. (\ref{eq:7}). 
$u$ and $v$ are the BCS occupation factors. 
${{\cal D}^{(i,f)}_{pn}}$ is the renormalized factor
of the RQRPA \cite{}simn96. In the case of the standard QRPA its value
is unity.

The one-body transition density leading to the final excited
$0^+_1$ two-phonon state obtained within the boson expansion
approach is given as 
\begin{eqnarray}
\langle 0^+_1 | \widetilde{ [c^+_{p}{\tilde{c}}_{n}]_{JM}}
| J^\pi M, m_f>  = 
 ( v_{p}^{(f)} u_{n}^{(f)} {\overline{X}}^{m_f}_{(pn, J^\pi)}
+ u_{p}^{(f)} v_{n}^{(f)} {\overline{Y}}^{m_f}_{(pn, J^\pi)} )
\left( {\cal D}_{pn} \right)^{-1/2}~\xi(p,p',n,n').
\label{eq:12}   
\end{eqnarray}
The factor $\xi(p,p',n,n')$ includes the details of the phonon
operator $\Gamma^{1\dagger}_2$ and can be found in Ref. \cite{sim01}.

\section{The light and heavy Majorana neutrino mass mechanisms}

The helicity flip Majorana neutrino mass mechanism is the most popular and
most commonly discussed $0\nu\beta\beta$-decay mechanism in the literature. 
Usually, the cases of the $0\nu\beta\beta$-decay mediated by 
light and heavy neutrinos are considered separately \cite{si99,bobi}. The relevant
lepton number violating parameters are
\begin{eqnarray}
\eta_{\nu}~ = ~\left|\frac{<m_\nu >_{e e}}{m_e}\right|^2, ~~~
\eta_{_N} ~ = ~\left|\sum^{heavy}_k~ (U_{e k})^2 ~ 
~\Xi_k ~ \frac{m_p}{M_k}\right|^2,
\label{eq:13}   
\end{eqnarray}
where the effective light electron neutrino mass is
\begin{equation}
<m_\nu >_{ee} ~= ~ \sum^{light}_k~ (U_{e k})^2 ~ \xi_k ~ m_k.
\label{eq.14}
\end{equation}
Here, $U_{e k}$ are the elements of the 
mixing matrix  for the electron neutrino
with the mass eigenstates  of the light $\chi_{k}$ and heavy $N_{k}$ Majorana
neutrinos with masses $m_k$ ($m_k \ll 1$ MeV) and $M_k$ ($M_k \gg 1$ GeV).
$\xi_k$ and $\Xi_k$ are the Majorana  phase factors.
$m_e$ and $m_p$ are the masses of the electron and the proton,
respectively.

It is worthwhile to notice that the light (heavy) neutrino exchange contribution
always leads to a long-range (short-range) neutrino potential inside the nucleus.
This has important implications for the evaluation of the
$0\nu\beta\beta$-decay nuclear matrix elements. 
The two-nucleon  light and heavy neutrino exchange 
potentials 
($H^{\eta_{\nu}}_{I}(r_{12})$ and $H^{\eta_{_N} }_{I}(r_{12})$, 
$I = F, ~GT, ~T$) can be written as \cite{si99}
\begin{eqnarray}
H^{\eta_{\nu}}_{I}(r_{12}) 
&=&\frac{2}{\pi g^2_A}
\frac{R}{r_{12}} \int_{0}^{\infty} 
\frac{\sin(qr_{12})} {q+E^m(J)- (E^i + E^f)/2} h_{I}(q^2)\,d{q}, 
\nonumber \\
H^{\eta_{_N} }_{I}(r_{12}) &=& \frac{1}{m_p m_e}\frac{2}{\pi g^2_A}
\frac{R}{r_{12}} \int_{0}^{\infty} 
{\sin(qr_{12})}h_{I}(q^2)~ q~ d{q} 
\label{eq:15}   
\end{eqnarray}
with
\begin{eqnarray}
h_{F}(q^2) &=& g^2_V (q^2) g^2_A, \nonumber \\
h_{GT}(q^2) & = & 
g^2_A (q^2) +\frac{1}{3}\frac{g^2_P (q^2) q^4}{4 m^2_p}
-\frac{2}{3}\frac{g_A (q^2) g_P (q^2) q^2}{2 m_p}
+\frac{2}{3} \frac{g^2_M ({\vec q}^{~2}) {\vec q}^{~2} }{4m^2_p}, \nonumber \\
h_{T}(q^2) & = & 
\frac{2}{3}\frac{g_A (q^2) g_P (q^2) q^2}{2 m_p}
- \frac{1}{3}\frac{g^2_P (q^2) q^4}{4 m^2_p}
+\frac{1}{3} \frac{g^2_M ({\vec q}^{~2}) {\vec q}^{~2}}{4m^2_p}.
\label{eq:16}   
\end{eqnarray}
Here, $R=r_0 A^{1/3}$ is the mean nuclear radius \cite{si99} with $r_0=1.1~fm$. 
$E^i$, $E^f$ and $E^{m}(J)$ are  the energies of the 
initial, final and intermediate nuclear states with angular momentum  $J$,
respectively.
The momentum dependence of the vector, weak magnetism, axial-vector and pseudoscalar
formfactors ($g_V(q^2)$, $g_M(q^2)$, $g_A(q^2)$ and $g_P(q^2)$) can be found
in Ref. \cite{si99}.

\begin{table}[t]
\caption{The kinematical factors $G_{01}$, nuclear matrix elements $\cal{M}_\kappa$
($\kappa~ = ~\eta_\nu,~\eta_{_N},~\eta_{\lambda}$) 
and  ratios $\xi_{kin}$, $\xi_{\cal{M}}$ and
$\xi_{full}$ [see Eq. (\ref{eq:3})]  associated with  
$0\nu\beta\beta$-decay transitions to the 
first excited  $0^+$ state and  to the $0^+$ ground state
calculated for A=76, 82, 100 and 136 nuclei. 
}
\label{table.1}
\begin{center}
\begin{tabular}{|l|ccccc|}\hline\hline
& mech. & $^{76}Ge$ &$^{82}Se$ & $^{100}Mo$  & $^{136}Xe$ \\ \hline
 & \multicolumn{5}{c}{ $0^+_{g.s.} \rightarrow 0^+_{g.s.}$ 
$0\nu\beta\beta$-decay transition} \\
 $E_i-E_f$ [MeV] &  & 3.067  &   4.027  &   4.055  &   3.503 \\
 $G_{01} $ [$y^{-1}$] & & $7.98\times 10^{-15}$ & $3.52\times 10^{-14}$ &
 $5.73\times 10^{-14}$ &  $5.92\times 10^{-14}$ \\
 $\cal{M}$     & $\eta_\nu $                & 2.80  & 2.64  & 3.21 & 0.66 \\
               & $\eta_{_N}$               & 32.6  & 30.0  & 29.7 & 14.1 \\
               & ${\eta_{\lambda}}$ & -625.  & -583.  & -750.  & -367. \\ 
 & \multicolumn{5}{c}{ $0^+_{g.s.} \rightarrow 0^+_{1}$ 
$0\nu\beta\beta$-decay transition} \\
 $E_i-E_f$ [MeV]  &  &  1.945  &   2.539 &    2.925 &    1.924 \\
 $G_{01} $ [$y^{-1}$] & &  $ 6.58\times 10^{-16}$ & $3.25\times 10^{-15}$ &  
 $1.11\times 10^{-14}$ & $2.81\times 10^{-15}$ \\
 $\cal{M}$     & $\eta_\nu $                & 0.994  & 0.947  & 1.76  & 0.441  \\
               & $\eta_{_N}$               & 16.3   &  15.2  & 16.2  & 10.5 \\
               & ${\eta{\lambda}}$ & -198.  & -185.  & -221.  & -136. \\ 
 & \multicolumn{5}{c}{ The suppression ratios} \\ 
 $\xi_{kin.}$  &     & 12.1  & 10.8  & 5.16  & 21.1  \\
 $\xi_{M.E.}$  & $\eta_\nu$                &  7.93 & 7.77  & 3.33  & 2.24 \\
               & $\eta_{_N}$               &  4.00 & 3.89  & 3.36  & 1.80 \\
               & ${\eta{\lambda}}$ &  9.96 & 9.93  & 11.5  & 7.28 \\ 
$\xi_{full}$   & $\eta_\nu$                &  96.  & 84.  & 17.  & 17. \\
               & $\eta_{_N}$               &  48.  & 42.  & 17.  & 38. \\
               & ${\eta{\lambda}}$ &  120. & 107. & 59.  & 153. \\ \hline\hline
\end{tabular}
\end{center}
\end{table}

\section{Trilinear R-parity breaking contribution}

Recently, the R-parity violation ($R_p \hspace{-1em}/\;\:$)
has been seriously considered in SUSY models.
It is defined as $R = (-1)^{3B+L+2S}$. B denotes the baryon number, L the lepton
number and S the spin of a particle leading to $R=1$ for SM particles and
$R=-1$ for superpartners. The lepton number violating part of the 
($R_p \hspace{-1em}/\;\:$) superpotential of the minimal supersymmetric SM can be written as  
\cite{Moh86,Ver87}
\begin{equation}
W_{R_{p}\hspace{-0.8em}/\;:}=\lambda _{ijk}L_{i}L_{j}E_{k}^{c}+\lambda
_{ijk}^{\prime }L_{i}Q_{j}D_{k}^{c} + \mu_i L_i H_2,  \label{W-Rp}
\end{equation}
where summation over flavor indices $i,~j, ~k$ is understood
($\lambda_{i,j,k}$ is antisymmetric in the indices $i$ and $j$).
$L$ and $Q$ denote lepton and quark doublet superfields 
while $E^{c},~D^{c}$ stand for
lepton and down quark singlet superfields.
Below we concentrate only on the trilinear $\lambda'$-couplings.

The trilinear $R_p \hspace{-1em}/\;\:$ SUSY mechanisms 
of the $0\nu\beta\beta$-decay 
are mediated by exchange of heavy SUSY particles, in particular
by squarks, gluinos and neutralinos \cite{Moh86,Ver87,FKSS97,awf99,kami}
 All these contributions 
involve only the ${\lambda'}_{111}$ coupling constant.
The half-life of the $0\nu\beta\beta$-decay
can be given in the form of Eq. (\ref{eq:3}) with the 
following lepton number violating parameter
\cite{FKSS97,awf99}):
\begin{equation}
\eta_{\lambda} =
\frac{\pi \alpha_s}{6}
\frac{\lambda^{'2}_{111}}{G_F^2 m_{\tilde d_R}^4}
\frac{m_p}{m_{\tilde g}}\left[
1 + \left(\frac{m_{\tilde d_R}}{m_{\tilde u_L}}\right)^2\right]^2.
\label{eq:17}
\end{equation}
Here, $G_F$ is the Fermi constant and
 $\alpha_s = g^2_3/(4\pi )$ is the $\rm SU(3)_c$ gauge coupling constant,
$m_{{\tilde u}_L}$, $m_{{\tilde d}_R}$ and
$m_{\tilde g}$  are masses of the u-squark, d-squark and
gluino. We assumed as usual the  dominance of the gluino exchange.

These SUSY mechanisms  of the $0\nu\beta\beta$-decay
were comprehensively investigated in a series of papers \cite{FKSS97,awf99}.
It turned out  that at the hadron level
the two-pion mode  $R_p \hspace{-1em}/\;\:$ SUSY contribution
to the $0\nu\beta\beta$-decay dominates over the conventional 
two-nucleon mode contribution. Due to this fact the nuclear matrix
elements associated with the heavy Majorana neutrino exchange
and $R_p \hspace{-1em}/\;\:$ SUSY mechanisms differ from 
each  other considerably. The Gamow-Teller and tensor 
parts of the two-nucleon potential  originating from the
pion-exchange $R_p \hspace{-1em}/\;\:$ SUSY mechanisms
are ($I~ =~GT,~T$)  
\begin{eqnarray}
H^{\eta_{\lambda}}_{I}(r_{12}) 
=
\left(\frac{m_A}{m_{_p}}\right)^2
\frac{m_{_p}}{ m_e}
      \left(
\frac{4}{3} \alpha^{1\pi} F^{(1)}_{I}(x) + 
\alpha^{2\pi} F^{(2)}_{I}(x) \right)
\label{eq:18}   
\end{eqnarray}
where 
\begin{eqnarray}
F_{GT}^{(1)}(x) &=&  e^{- x}, \ \ \ F_{T}^{(1)}(x) =
(3 + 3x + x^2)\frac{e^{- x}}{x^2},\\
F_{GT}^{(2)}(x) &=& (x - 2) e^{- x}, \ \ \ F_{T}^{(2)}(x) = (x + 1) e^{- x},~~
x = m_\pi r_{12}.
\label{eq:19}
\end{eqnarray}
Here, $m_A (= 850~MeV)$ is the cutoff mass of nucleon formfactor
and $m_\pi$   the mass of the pion.
Values of the structure coefficients $\alpha^{k\pi}$ (k=1,2)
are \cite{FKSS97}: $\alpha^{1\pi} = -4.4\times 10^{-2}$,   
$\alpha^{2\pi} = 2\times 10^{-1}$.  We  note that the Fermi part of the
two-nucleon exchange potential is equal to zero for the above discussed
mechanism.   

\section{ Discussions and Conclusions}

\begin{figure}[tb]
\epsfysize=15.0cm
\begin{center}
\epsfig{file=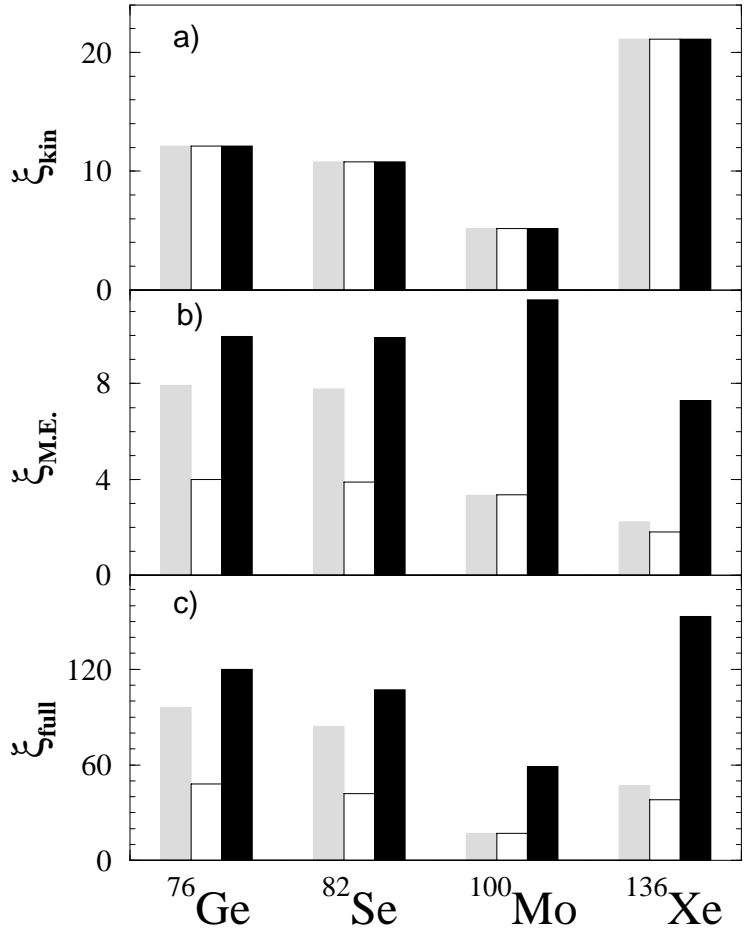,scale=0.8}
\end{center}
\caption{The suppression ratios $\xi_{kin}$ (a), $\xi_{\cal{M}}$ (b) and
$\xi_{full}$ (c)  for  transition to the first
excited $0^+$ state in the final nucleus relative the transition to the $0^+$ ground state
for A=76, 82, 100 and 136 nuclei. For the definition
of these factors see Eq. (\ref{eq:3}).
The opaque, open and black bars  correspond to results for the
light Majorana neutrino mass, heavy Majorana neutrino mass and 
R-parity breaking SUSY mechanisms, respectively.
\label{fig.2}}
\end{figure}

In this contribution we calculated  ratios 
$\xi_{kin}$, $\xi_{\cal{M}}$ and
$\xi_{full}$ [see Eq. (\ref{eq:3})]  for the 
$0\nu\beta\beta$-decay to the
ground state $0^+_{g.s.}$ 
and the first excited $0^+_1$ state
for the decays 
$^{76}Ge\rightarrow {^{76}Se}$,   
$^{82}Se\rightarrow {^{82}Kr}$,   
$^{100}Mo\rightarrow {^{100}Mo}$  and 
$^{136}Xe\rightarrow {^{136}Ba}$.  
Their knowledge is important for distinguishing among 
different $0\nu\beta\beta$-decay mechanisms.
We studied the light
and heavy Majorana neutrino mass and
the trilinear R-parity breaking mechanisms
of the $0\nu\beta\beta$-decay. 
The  results are summarized in Table \ref{table.1}
and displayed in Fig \ref{fig.2}.  
For the sake  of completeness we repeat the  results  
of Ref. \cite{sim01} 
for the kinematical factors and the nuclear matrix elements
in Table \ref{table.1} as well.

In Table \ref{table.1} one sees that 
the values of $\xi_{kin}$ are 12.1, 10.8, 5.16 and
21.1 for A= 76, 82, 100 and 136 systems, respectively.
Partially by these factors, which are not related to the
specific $0\nu\beta\beta$-decay mechanism, the decays
to the excited $0^+_1$ states are suppressed in comparison 
with decays to the $0^+$ ground states. The additional 
suppression comes from the smaller values of the nuclear 
matrix elements in the case of the  
$0\nu\beta\beta$-decay to the $0^+_1$ excited states in 
comparison with those to $0^+$ ground states.
We remind that nuclear matrix elements to the first
$0^+$ excited final states were evaluated within a 
boson expansion approach. The reliability of
results obtained and their comparison with
results of other approaches have been studied in
Ref. \cite{sim01}. As it was already mentioned 
before, the factor $\xi_{\cal{M}}$ depends on the 
mechanism considered. 
It follows from  Fig. \ref{fig.2} that $\xi_{\cal{M}}$  
is largest in the case of the $R_p \hspace{-1em}/\;\:$
SUSY mechanism. The largest difference between
$\xi_{\cal{M}}$ for the different mechanisms
is found in   A = 100
and 136 nuclei.  Fig. \ref{fig.2}  implies
that the suitable candidates to distinguish between
light and heavy Majorana neutrino mechanism
are the A = 76 and 82 nuclei. We note that the behavior
of the ratios $\xi_{full}$ corresponds to  
factors $\xi_{\cal{M}}$. 
The value of $\xi_{full}$ gives the suppression
of the decay to the excited $0^+_1$ state in
comparison the decay to the $0^+_{g.s.}$ ground state
for a given $0\nu\beta\beta$-decay mechanism. 

We thus reach a very important conclusion. 
It is possible to distinguish among the light and heavy
Majorana neutrino mass and $R_p \hspace{-1em}/\;\:$
SUSY breaking mechanisms of the $0\nu\beta\beta$-decay 
by studying the transitions to the first excited
$0^+_1$ states both theoretically and experimentally.  
The key issue in the theoretical analysis of these
$0\nu\beta\beta$-decay transitions is the reliable 
evaluation of the associated nuclear matrix elements
within the most advanced nuclear structure approaches.
There is  hope that the current
nuclear structure uncertainties can be reduced by further 
development of the nuclear models.
The experimental study of $0\nu\beta\beta$-decay 
to first excited $0^+_1$ state is  more challenging
as the observation of the ground state  transition. 
Since one has to 
improve the sensitivity of  $0\nu\beta\beta$-decay
experiments to observe the decay to the first
$0^+$ excited final state with half-lifes by a factor
of 10-100 larger in comparison with the half-lifes 
for the ground state to ground state transition.    
At least it is also important to collect
experimental information from more
$0\nu\beta\beta$-decay isotopes. It could
help substantially to reduce the chances of 
wrong conclusions. We note
that many ambitious projects 
(NEMO III, CUORE, GENIUS, GEM, MAJORANA, MOON, XMASS etc.
\cite{avi,cuore,nemo3,Genius,GEM,ejiri,Borex}) 
are under way. Perhaps, some of them have the chance to
observe both $0\nu\beta\beta$-decaies 
to the $0^+$ ground and the first excited states.

\vspace{0.5cm}

{\bf Acknowledgements:} This work was supported by  the Deutsche 
Forschungsgemeinschaft under contract 436 SLK 17/298.


\end{document}